\documentclass[aps,prl,twocolumn,superscriptaddress]{revtex4}
\usepackage[dvips]{graphics}

\begin{document}
 \title{Marangoni shocks in unobstructed soap-film flows}
\author{Tuan Tran}
\affiliation{Department of Mechanical Science and Engineering, University of Illinois, Urbana, IL 61801}
\author{Pinaki Chakraborty}
\affiliation{Department of Geology, University of Illinois, Urbana, IL 61801}
\author{Gustavo Gioia}
\affiliation{Department of Mechanical Science and Engineering, University of Illinois, Urbana, IL 61801}
\author{ Stanley Steers}
\author{Walter Goldburg}
\affiliation{Department of Physics and Astronomy, University of Pittsburgh, Pittsburgh, PA 15260 }

\date{\today}

\begin{abstract}
It is widely thought that in steady,
gravity-driven, unobstructed soap-film
flows, the velocity increases monotonically
downstream. Here we show experimentally that
the velocity increases, peaks, drops abruptly,
then lessens gradually downstream. We argue
theoretically and verify experimentally that
the abrupt drop in velocity corresponds to
a Marangoni shock, a type of shock
related to the elasticity of the film.
Marangoni shocks induce locally intense turbulent
fluctuations and may 
  help elucidate the mechanisms that produce
 two-dimensional turbulence
away from boundaries.
\end{abstract}

\maketitle


 Soap-film flows  \cite{soapf} 
 have long been used to study two-dimensional (2-D) 
 turbulence, a type of turbulence that differs from its 
 three-dimensional (3-D) counterpart in crucial respects.
 For example, in 3-D turbulence the energy 
 may cascade only from larger to smaller lengthscales
 whereas in 2-D turbulence the 
 energy may cascade in either 
  direction  \cite{cascade}. 
 Disparate directions of
 energy transfer result
 in disparate apportionings of the 
  turbulent kinetic energy among the lengthscales of the flow 
 \cite{cascade}. 
 Besides the theoretical interest 
 inherent in its distinctive characteristics, 
 2-D turbulence is relevant to the large-scale 
 irregularities  encountered in 
 2-D atmospheric flows---flows that
  are confined to two dimensions 
  by geostrophic forces and  a stratified 
 atmosphere \cite{twod}. 
 Examples of large-scale 
 irregularities in 2-D atmospheric flows include 
 hurricanes, typhoons, 
 and the Great Red Spot
 of Jupiter \cite{atmo}.

\begin{figure} 
\centering
\resizebox{1.3in}{!}{\includegraphics{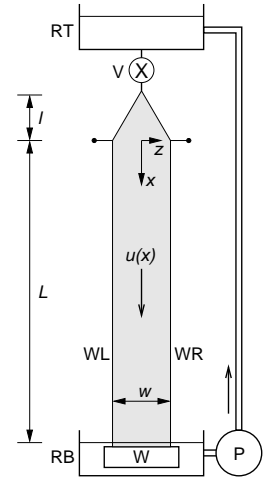}}
\caption{ \label{setup}
 Typical 
 setup used to study
 steady, gravity-driven, unobstructed
  soap-film flows. Axis $x$  
 runs vertically along the centerline of the flow.
 Wires WL and WR are thin nylon-fishing lines kept taut by  
 weight W. The film hangs from the wires;
 its width increases from 0 to $w$ over
 an expansion section of length $l$,
 then remains constant and equal to $w$ over
 a measurement section of length $L\gg l$.
(The origin of $x$ is at the top of the
 measurement section. $L$ is the ``length of the flow''
 and $w$ the ``width of the flow.'')
 Reservoir RT contains a soapy solution
 which flows through
  valve V and into the film.
 After flowing through the film 
  with mean velocity $u(x)$, 
 the soapy solution drains into reservoir RB 
  and returns to reservoir RT via pump P.
 In our experiments, the soapy solution
 consists of $\simeq 2.5\,$\% Dawn Nonultra 
 in water; $w= 2.5$ to 5.1$\,$cm;
  $L=1.05$ to 1.39$\,$m;
 and  $l=23.5\,$cm.}
\end{figure}

In the typical 
 setup used to 
  study soap-film flows  \cite{rutgers2001}, a film 
 hangs between two long, vertical, mutually parallel
  wires a few centimeters apart from one another. 
  Driven by gravity, a steady vertical flow soon 
 becomes established within the film  (Fig.~\ref{setup}).
 Then, the thickness $h$ of the film is roughly
 uniform on any 
 cross section of the 
 flow \cite{georgiev2002}, and we 
  write $h=h(x)$, where $x$ 
 runs along the centerline of the flow
 (Fig.~\ref{setup}).  In a typical flow
  $h\approx 10\,\mu$m, much smaller
 than both the width $w$ and the length $L$
 of the flow  (Fig.~\ref{setup}). As a result,
 the velocity of the flow 
 lies on the plane of the film, and 
 the flow is 2-D. Since the viscous stresses
  (and the attendant velocity gradients) 
 are confined close 
 to the wires, the mean (time-averaged)
 velocity  $u$ is roughly uniform on any 
  cross section of the film  \cite{rutgers1996},
  and we write $u=u(x)$. 
 Thus, assuming 
  incompressibility, 
 $h(x) u(x)$ 
 equals the flux $q$ per unit width of film
 and is independent of $x$ for a steady flow.
 
 Analyses of steady flows have accounted for 
 the gravitational force, 
 the inertial force, the drag force of the ambient air,
  and the drag force of the wires.
  Rutgers et~al.\  \cite{rutgers1996} have shown that
 the drag force of the wires is negligible 
    as compared to the drag force of the 
 ambient air and may be dropped from the 
 equation of momentum balance.
 Then, a prediction can be made that in a steady flow
   the mean velocity 
 is a monotonically increasing function of $x$
  and approaches a terminal velocity asymptotically
 downstream \cite{rutgers1996, georgiev2002}. 
 This prediction has not been tested,
  but it is thought to be 
 in qualitative agreement with the few known
  experiments 
  \cite{rutgers1996, georgiev2002}.  
In contrast to this prediction,
  in our experiments $u(x)$ 
is a strongly non-monotonic function of $x$.

\begin{figure}[htbp]
\centering
\resizebox{2.55in}{!}{\includegraphics{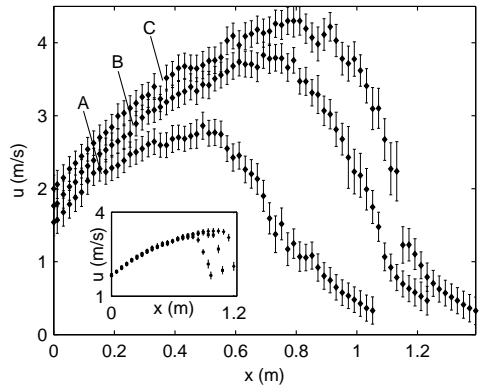}}
\caption{\label{velo}
Plots of the mean velocity $u$ vs.\ $x$ 
  for three steady flows. The width $w=5.1\,$cm
 for all flows; both the length $L$ and the flux
 per unit width $q$ change from flow to flow.
  $L=1.05\,$m and 
   $q= 3.8 \, 10^{-6}\,$m$^2$/s (A),
   $L=1.23\,$m  and $q= 4.9 \, 10^{-6}\,$m$^2$/s (B),
   $L=1.39\,$m and $q= 5.5 \, 10^{-6}\,$m$^2$/s (C).
  Inset: Plots of $u$ vs.\ $x$ 
  for three steady flows. $w=5.1\,$cm and
  $q= 5.7 \, 10^{-6}\,$m$^2$/s for all flows;
  $L$ changes from flow to flow. 
    }
\end{figure}

To measure $u$, we use
 laser Doppler velocimetry (LDV).
 In Fig.~\ref{velo} we show plots 
 of $u$  along the centerline of several representative flows.
  In each flow, $u$ increases downstream
  up to a certain point whereupon it peaks,
  drops abruptly to a fraction of its peak value, then 
 continues to lessen gradually downstream.

\begin{figure}[htbp]
\centering
\resizebox{2.7in}{!}{\includegraphics{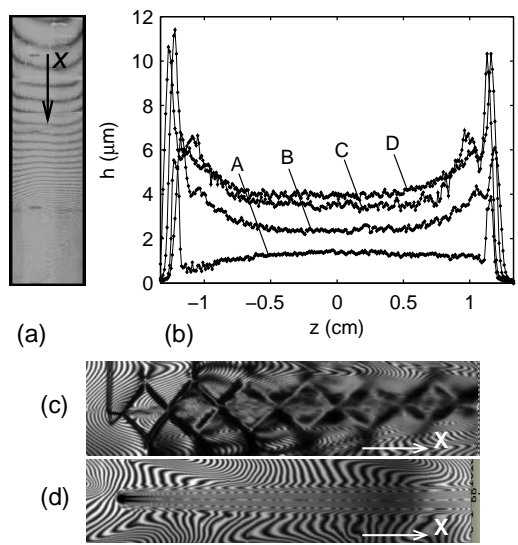}}
\caption{\label{thick}
(a)  Fringes over the part of a
  film where the velocity drops abruptly. 
 (b) Plots of the thickness vs.\ $z$ 
 along four 
  cross sections of a 
 flow of width $2.5\,$cm and length $1.2\,$m.
 The cross sections are
 at $x=0.95\,$m (A),  $x=1.04\,$m (B), $x=1.05\,$m (C),
 and $x=1.07\,$m (D). 
  The large peaks near the lateral 
 edges 
   are due to backscattering from the wires.
 Fringes at a piercing upstream (c) and downstream (d) 
 of the drop in velocity; fields of view $=5\,$cm$\times 1\,$cm.
}
\end{figure}

 From the incompressibility condition ($uh=q$),
the abrupt drop in $u$ should
  be accompanied
 by an abrupt increase in $h$. 
To verify this
  abrupt increase in $h$, 
 we light one face
 of a film with a 
   sodium lamp
 and observe the interference fringes 
that form there.
In Fig.~\ref{thick}a we show 
 a photograph of the interference
  fringes on the part of a 
  film where $u$ drops abruptly. 
  The distance between 
 successive fringes decreases
   rapidly in the downstream direction, 
 signaling an abrupt increase in $h$. 

To verify the 
 abrupt increase in $h$ by means of an alternative
 technique,  we put Flourescein dye in the soapy
solution and
focus incoherent blue light on
  a spot on  the film.  The
 spot becomes fluorescent, and we monitor 
 the intensity of the fluorescence using
  a photodetector whose counting 
 rate is proportional to $h$.
 In Fig.~\ref{thick}b we show plots
 of $h$ along four 
 cross sections
  of a flow. These  cross sections
   lie on the part of the flow
  where $u$ drops abruptly. 
  The thickness trebles
 over a short
  distance of a few centimeters
  in the downstream direction.

To explain our experimental results, we 
 write the steady-state equation of momentum balance in the form  
\begin{equation} \label{start}
	\rho h u u_x= 2 \sigma_x +\rho g h  - 2 \tau_a,
\end{equation} 
where   $\rho$ is the density,
 $(\cdot)_x=d(\cdot)/dx$, $\sigma$ is the surface tension,
 $g$ is the gravitational acceleration,
 and $\tau_a$ is the shear stress 
 due to air friction. From left to right, the
 terms in (\ref{start}) represent the inertial force,
  the elastic force, the gravitational force, 
 and the drag force of the ambient air.
 Here we follow Rutgers et~al.\ 
 \cite{rutgers1996} and use (as a rough approximation)
  $\tau_a = 0.3\sqrt{\rho_a \mu_a u^3/(x+l)}$, the 
 Blasius expression 
 for the  shear stress on a rigid plate that moves at
  a constant velocity $u$ through air 
 of density $\rho_a=1.2\,$kg/m$^3$  
 and viscosity $\mu_a=1.7\times10^{-5}\,$kg/ms.

 To obtain an expression for $\sigma_x$, we argue 
 that  the concentration of 
 soap molecules in the bulk of the film remains 
 constant  in  our experiments (because there is no time 
  for diffusional exchange between the bulk and the
  faces of the film \cite{couder1989,maratimeNEW}). 
 Then,  the film is said to be in the 
 \emph{Marangoni regime},
 and $2\sigma_x=-\rho U_M^2 h_x$ \cite{mararegNEW},
 where $U_M$ is the \emph{Marangoni speed}---a 
 property of the film, independent of $h$, that quantifies
  the speed at which disturbances in $h$
  travel on the plane of the film \cite{couder1989,mararegNEW}.
    By substituting $2\sigma_x=-\rho U_M^2 h_x$ and $h=q/u$ 
  in (\ref{start}), 
  we obtain the governing equation
\begin{equation} \label{goveq}
	u_x=u \frac {g-2\tau_a u/\rho q}{u^2-U_M^2}.
\end{equation}

In (\ref{goveq}) we can distinguish two 
  types of flow: a supercritical flow
  in which $u>U_M$ and $u_x>0$,  
 and a subcritical flow in which $u<U_M$ and $u_x<0$. 
 We conjecture that in 
  our experiments the flow is supercritical 
 upstream of the drop in velocity and subcritical 
 downstream. Consistent 
 with this conjecture, for any fixed $q$ the flow 
upstream of the 
 drop in velocity remains invariant to 
 changes in the length of the flow  (inset of Fig.~\ref{velo}).
  
To confirm that flows are supercritical 
 upstream of the drop in velocity and subcritical 
 downstream, we use pins
 to pierce a flow upstream and downstream 
 of the drop in velocity (Figs.~\ref{thick}c and d, respectively).
  Upstream of the drop in velocity the 
  Mach angle $\approx 50^o$ (from Fig.~\ref{thick}c),
 and the local $u=1.83\,$m/s 
 (from an LDV measurement); thus 
  $U_M\approx\sin50^o \times  1.83\,$m/s$ =1.4\,$m/s
  in our experiments.

\begin{figure}[htbp]
\centering
\resizebox{2.7in}{!}{\includegraphics{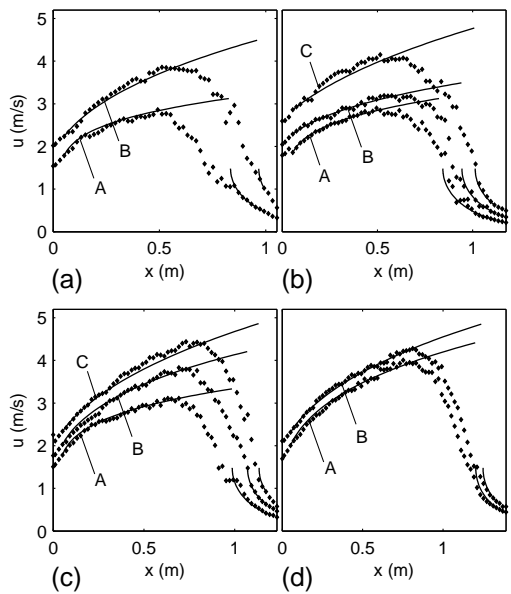}}
\caption{\label{compa}
 Plots of the computational $u(x)$ (lines)
 and  experimental $u(x)$
 (points) for ten different flows.
 The computations are for $U_M= 1.48\,$m/s and 
 the values of $q$ indicated below
 (the experimental estimates for $q$ are
  indicated in parentheses \cite{flux}). 
    $w=5.1\,$cm for all flows.
   (a) Flows of length $1.05\,$m:
 (A) $q= 5.7\, 10^{-6}\,$m$^2$/s 
 ($3.9 \, 10^{-6}\,$m$^2$/s),
   (B) $25 \, 10^{-6}\,$m$^2$/s 
 ($5.7 \, 10^{-6}\,$m$^2$/s);   
 (b) flows of length $1.17\,$m:
 (A) $5.7 \, 10^{-6}\,$m$^2$/s
($5.1 \, 10^{-6}\,$m$^2$/s),
  (B) $7.4 \, 10^{-6}\,$m$^2$/s 
 ($5.9 \, 10^{-6}\,$m$^2$/s), 
  (C) $30 \, 10^{-6}\,$m$^2$/s 
 ($7.5 \, 10^{-6}\,$m$^2$/s); 
 (c) flows of length $1.23\,$m:
 (A) $6.1 \, 10^{-6}\,$m$^2$/s
 ($4.1 \, 10^{-6}\,$m$^2$/s),
 (B)  $14 \, 10^{-6}\,$m$^2$/s
 ($5.3 \, 10^{-6}\,$m$^2$/s), 
 (C) $31 \, 10^{-6}\,$m$^2$/s
 ($6.5 \, 10^{-6}\,$m$^2$/s); 
and (d)  flows of length $1.39\,$m:
 (A) $16 \, 10^{-6}\,$m$^2$/s 
 ($4.7 \, 10^{-6}\,$m$^2$/s), 
  (B) $25 \, 10^{-6}\,$m$^2$/s
 ($6.3 \, 10^{-6}\,$m$^2$/s). }
\end{figure}

  Let us test the governing equation (\ref{goveq})
  for one of our experiments. We adopt a value of $U_M$ and
 a value of $q$ and perform two computations.
First, we integrate 
 (\ref{goveq})
  downstream from $x=0$ 
 with boundary condition $u(0)=u_0$, where $u_0$
 is the velocity measured at $x=0$
 in the experiment \cite{bcond}. 
 This first computation gives a function $u(x)$ that 
 should fit the experiment upstream of the drop in velocity.
  Second, we integrate 
 (\ref{goveq}) upstream from 
 $x=L$ with boundary condition $u(L)=u_{L}$, 
 where $u_{L}$ is the velocity measured at $x=L$
  in the experiment \cite{bcond}.  
This second computation
  gives a function $u(x)$ that 
should fit the experiment
  downstream of the drop in velocity. 
 We perform the same computations for each one of our
  experiments trying different values of 
 $q$ and $U_M$,  and choose the 
 \emph{optimal values of} $q$   and the 
  \emph{optimal value of} $U_M$---that is to say, 
 the values of $q$ (one for each experiment) 
 and the value of $U_M$  (the same for all experiments)
  that yield the best fits 
   to the experiments (Fig.~\ref{compa}).
 The optimal value of $U_M$, $1.48\,$m/s, is
 in remarkable agreement 
 with our 
 estimate  from Fig.~\ref{thick}c ($1.4\,$m/s).
 The optimal values of $q$ 
are in reasonable agreement with 
 the experimental estimates for $q$ 
  \cite{flux} (caption to Fig.~\ref{compa}).

We conclude that
  a drop in velocity 
  signals a supercritical-to-subcritical transition 
  and corresponds to 
 a \emph{Marangoni shock}. 
  In theory the drop in velocity
 is infinitely steep and may be said to
 take place at $x=x^*$,
  where $u$ attains the
   value of $U_M$ (and $u_x$ becomes singular) 
  in the subcritical flow  (Fig.~\ref{compa})
  \cite{theojump}. But 
  in experiments the drop in velocity
  takes place over a finite span $\Delta x$ whose 
  magnitude appears to increase
  with $q$  (Fig.~\ref{compa}) and  whose 
 downstream edge is located 
 at about $x=x^*$, the theoretical position of the 
 shock (a position which appears to move 
 downstream as $q$ increases). Thus in 
 our simple theory 
 the shock is sharp whereas in experiments 
  the shock is diffused over a finite 
  span $\Delta x$.

 To understand the reason why our theory (which does
 not account for turbulence) cannot resolve 
 the structure of the shock, 
  recall that a shock must dissipate energy 
 at a steady rate \cite{power}. 
 We argue (i) that the shock dissipates energy by 
 powering locally intense
  turbulent fluctuations and (ii) that 
 these fluctuations must extend roughly over
  the same span $\Delta x$ as the 
  shock that powers them \cite{faber1997}.
    
\begin{figure}[htbp]
\centering
\resizebox{2.5in}{!}{\includegraphics{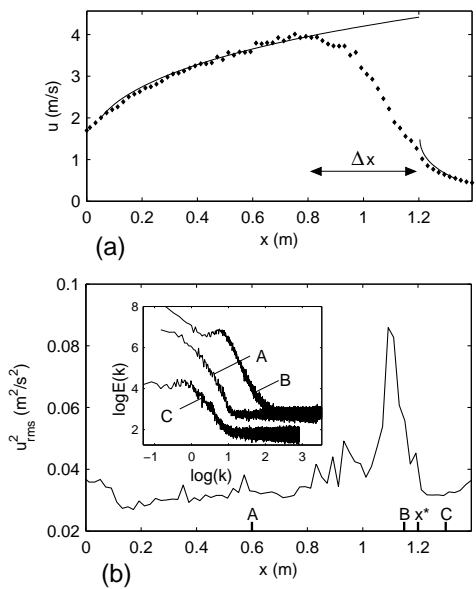}}
\caption{\label{rms}
(a)  Plots of the computational $u(x)$ (lines)
 and the experimental $u(x)$ 
 (points) for a representative flow.
 $\Delta x$ is the span of the shock.
 (b) Plot of the experimental $u_{\rm rms}^2$
 (an index of the energetic contents of the
 velocity fluctuations)
  vs.\ $x$ for the same flow. Inset: 
  energy spectra at the centerline
 of the same flow for the cross section at 
 $x=0.60\,$m (A),
$x=1.17\,$m (B), and
$x=1.30\,$m (C). These are the cross sections 
 marked ``A,'' ``B,'' and ``C'' in part b. 
 The spectra are log-log
  plots of the energy density
 $E$ (m$^3$/s$^2$) vs.\ the wavenumber $k$ (1/m).
}
\end{figure}

   To test these arguments we use LDV
 to measure 
 the root-mean-square velocity
 $u_{\rm rms}$ along the centerline 
 of a representative flow (Fig.~\ref{rms}).
  From a comparison of Figs.~\ref{rms}a
 and b, we confirm that the  
 shock is accompanied
  over its entire span
  $\Delta x$  by velocity fluctuations
   that are up to thrice as intense as the 
 velocity fluctuations that prevail 
  both upstream and dowstream of $\Delta x$. 
  We conjecture that
intense velocity 
  fluctuations can arise more readily  
  where the mean velocity is higher; 
 this 
 may explain why the locally intense velocity 
  fluctuations---and the
 diffusive shock that powers them---are 
 located on the supercritical side of the theoretical
 position of the shock. 

To verify that the velocity fluctuations 
  are turbulent, we obtain the energy 
 spectrum on the centerline of the flow
 for three cross sections: one upstream,
 one within, and one  
  downstream of the shock (inset to 
  Fig.~\ref{rms}b).
 (These cross sections are marked ``A,'' ``B,''
 and ``C'' in Fig.~\ref{rms}b.)
 The area under the 
 spectrum is 
  larger for cross section B 
than for 
cross sections A and C, confirming
 that the turbulence is more intense
  within the shock than elsewhere
 in the flow. 
 Further, the slope of the spectrum at intermediate
  wavenumbers and the shape of the spectrum at low wavenumbers
  differ on either side of the shock, indicating
 that the spectrum undergoes
 structural changes as the flow traverses the shock.  
  
We have demonstrated 
 the spontaneous occurrence of shocks in the  
 soap-film flows that are customarily used
 in experimental work on two-dimensional turbulence.
 These shocks are dissipative and diffusive;
 they give rise to fluctuations independently from
  the boundaries, with
 a strong but circumscribed effect
 on the spatial distribution of turbulent intensity; 
and they
  alter the structure
  of the turbulent spectrum downstream from the 
 shock. We conclude that the presence 
 of shocks should be factored in in the
 interpretation of experimental measurements,
 and submit that shocks may furnish a convenient setting
  to study localized turbulence production
 in two dimensions.

\section{Acknowledgements}  
  NSF funded this reasearch through
  grants DMR06--04477 (UP)
 and DMR06--04435 (UIUC). 
 The Vietnam Education Foundation funded T.\ Tran's work.
We thank Alisia Prescott, Jason Larkin, Nik Hartman, Hamid Kellay,
  Nicholas Guttenberg,  and Nigel Goldenfeld.


\begin{thebibliography}{} 
\bibitem{soapf} M.\ Gharib and P.\ Derango, Physica D {\bf 37}, 406 (1989);
 M.\ Beizaie and M.\ Gharib, Exp.\ Fluids {\bf 23}, 130 (1997); H.\ Kellay and W.\ Goldburg, Rep.\ Prog.\ Phys.\ {\bf 65}, 845 (2002);
 P.\ Tabeling, 
 Phys.\ Rep.\ {\bf 362},  1 
  (2002).

\bibitem{cascade} R.\ Kraichnan, Phys.\ Fluids {\bf 10}, 1417 (1967); 
   G.\ Batchelor, ibid.\  {\bf 12}, II-233 (1969);  
 R.\ Kraichnan and D.\ Montgomery, Rep.\ Prog.\ Phys.\ {\bf 43}, 547 (1980).

\bibitem{twod} J.\ Pedlosky, {\it Geophysical Fluid Dynamics\/} 
 (Springer-Verlag, 1987). 

\bibitem{atmo}  C.\ Leith, J.\ Atmos.\ Sci.\ {\bf 28}, 145 (1971);
  C.\ Leith and  R.\ Kraichnan, ibid.\ {\bf 29}, 1041 (1972);
  M.\ Lesieur, {\it Turbulence in Fluids\/} 
 (Kluwer Academic Publishers, Dordrecht, 1997);
  P.\ Marcus, Nature {\bf 428}, 828 (2004);
  F.\ Seychelles et al., Phys.\ Rev.\ Lett.\ {\bf 100}, 144501 (2008). 

\bibitem{rutgers2001} M.\ Rutgers, X.\ Wu and W.B.\ Daniel, 
 Rev.\ Sci.\ Instrum.\ {\bf 72}, 3025 (2001).

\bibitem{georgiev2002}
 D.\ Georgiev and P.\ Vorobieff, Rev.\ Sci.\ Instrum.\ {\bf 73}, 1177 (2002).


\bibitem{rutgers1996} M.\ Rutgers et al., Phys.\ Fluids {\bf 8}, 2847 (1996).

\bibitem{couder1989}Y.\ Couder, J.M.\ Chomaz, and M.\ Rabaud,
  Physica D {\bf 37}, 384 (1989);
  C.Y.\ Wen, S.K.\ Chang-Jian, and M.C.\ Chuang, 
 Exp.\ Fluids {\bf 34}, 173 (2003).

\bibitem{maratimeNEW}
Note that a change in $h$ 
 (and the attendant stretching of the film) may disturb the mutual 
 equilibrium between the bulk and the faces of the film.
 To show that the concentration of soap molecules in the 
  bulk remains constant,
  we must show that the rate of change of $h$ does not allow time for
  soap molecules to diffuse between the bulk and the
  faces of the film, so that  $t_D$ (the timescale  of diffusion) $\ge t_d$ 
 (the timescale associated with changes in $h$). 
  For $h=10\,\mu$m (a typical value
  in our experiments) we estimate $t_D = 1\,$s \cite{couder1989}.
  To obtain an upper bound on $t_d$, we argue that $t_r$ 
 (the residence time of a drop of soapy solution in the flow)
  $\gg t_d$.
  For $u=1\,$m/s and $L=1\,$m (typical values in our experiments)
  we estimate $t_r = L/u =
  1\,{\rm s}$, and conclude that $t_D \approx t_r \gg t_d$.

\bibitem{mararegNEW}
The surface tension of a dilute soap solution can be expressed
  as $\sigma=\sigma_0 - RT\Gamma$, where $\sigma_0$ is the surface tension of
  pure water, $R$ is the gas constant, $T$ is the absolute temperature, and
  $\Gamma$ is the concentration of soap molecules on the faces of the 
 film \cite{couder1989}. Now, by definition $\Gamma=(c-c_b)h/2$, where 
  $c$  is the overall concentration of soap molecules in the film
  and $c_b$ is the concentration of soap molecules 
  in the bulk of the film. 
  Since $c$ remains constant (because of incompressibility) and 
  $c_b$ also remains constant (because the film is in the Marangoni regime),
  $\Gamma_x= (c-c_b)h_x/2$ and $\sigma_x=-\rho U_M^2h_x/2$, where 
   $U_M\equiv \sqrt{RT(c-c_b)/\rho}$ is a constant 
  independent of $h$.

\bibitem{bcond} In actuality, we integrate downstream from
  $x=w$ with boundary condition $u(w)=u_{w}$, where $u_{w}$
 is the velocity measured at $x=w$ in the experiment,
  and $w$ is the width of the flow ($5.1\,$cm). 
   Thus we avoid using the velocity measured 
  at $x=0$, where the flow is likely to be 
  disturbed by end effects associated with 
  the expanding section (Fig.~\ref{setup}).
 In an analogous way, we later integrate upstream from
  $x=L-w$ with boundary condition $u(L-w)=u_{L-w}$, where $u_{L-w}$
 is the velocity measured at $x=L-w$ in the experiment.

\bibitem{flux}  We estimate these values of $q$ 
 by measuring the volume of soapy solution
 that drains into reservoir RB (Fig.~\ref{setup})
 in a given time interval and divide this volume
by $w$.

\bibitem{theojump} It may be argued
 theoretically that the sharp shock
  is located downstream of $x^*$,
  where $u_+ u_- =U_M^2$ 
 (Rayleigh's jump condition, where $+$ and $-$ denote 
  down and upstream of the jump, respectively.).

\bibitem{power}
The energy per unit area on the plane of the film
 is the sum of the elastic,
 kinetic, and potential energy,
 $e=2 \sigma+\rho h u^2/2-\rho g h x$.
 Thus the energy conveyed per unit 
 time and unit width of film is $e u$, and the shock 
 must dissipate a power per unit width
  $P=-[\![ eu ]\!]$, 
where $[\![(\cdot)]\!] \equiv (\cdot)_+- (\cdot)_-$.
By substituting 
 $\sigma=\sigma_0-\rho U_M^2 h/2$
  \cite{mararegNEW} and $h=q/u$, we obtain 
  $P=-2 \sigma_0 [\![ u ]\!]-\rho q [\![ u^2 ]\!]/2>0$.

\bibitem{faber1997} Here we argue by analogy with turbulent 
 hydraulic jumps in open channels; see, e.g.
   T.E.\ Faber, {\it Fluid Dynamics for Physicists\/} 
 (Cambridge U.\ Press, 1997) and D.\ Bonn, A.\ Anderson, and
 T.\ Bohr, 
 J.\ Fluid Mech.\ {\bf 618},  71 
 (2009). 
\end{thebibliography}
\end{document}